\documentclass[aps,prl,twocolumn,floatfix,superscriptaddress,showpacs,amsfonts,amssymb,amsmath,preprintnumbers]{revtex4}
\usepackage{graphicx}

\newcommand\Alfven{Alfv\'en }
\newcommand{\V}[1]{\mathbf{#1}} 
 
\newcommand{\figref}[1]{Fig.~\ref{#1}}

\begin{document}


\title{Kinetic Simulations of Magnetized Turbulence in Astrophysical Plasmas}


\author{G.~G.~Howes}
\email[]{ghowes@astro.berkeley.edu}
\affiliation{Department of Astronomy, University of California, Berkeley, 
California 94720, USA.}
\author{W.~Dorland}
\affiliation{Department of Physics, CSCAMM, and IREAP, University of Maryland, College Park, 
Maryland 20742-3511, USA.}
\author{S.~C.~Cowley}
\affiliation{Department of Physics and Astronomy, UCLA, Los Angeles, California 90095-1547, USA.}
\affiliation{Plasma Physics, Blackett Laboratory, Imperial College, London SW7 2AZ, UK.}
\author{G.~W.~Hammett}
\affiliation{Princeton Plasma Physics Laboratory, Princeton, NJ 08543, USA.}
\author{E.~Quataert}
\affiliation{Department of Astronomy, University of California, Berkeley, 
California 94720, USA.}
\author{A.~A.~Schekochihin}
\affiliation{Plasma Physics, Blackett Laboratory, Imperial College, London SW7 2AZ, UK.}
\author{T.~Tatsuno}
\affiliation{Department of Physics, University of Maryland, College Park, 
Maryland 20742-3511, USA.}

\date{\today}

\begin{abstract}

This letter presents the first \emph{ab initio}, fully
electromagnetic, kinetic simulations of magnetized turbulence in a
homogeneous, weakly collisional plasma at the scale of the ion Larmor
radius (ion gyroscale). Magnetic and electric-field energy spectra
show a break at the ion gyroscale; the spectral slopes are consistent
with scaling predictions for critically balanced turbulence of \Alfven
waves above the ion gyroscale (spectral index $-5/3$) and of kinetic
\Alfven waves below the ion gyroscale (spectral indices of $-7/3$ for
magnetic and $-1/3$ for electric fluctuations). This behavior is also
qualitatively consistent with \emph{in situ} measurements of
turbulence in the solar wind.  Our findings support the hypothesis
that the frequencies of turbulent fluctuations in the solar wind
remain well below the ion cyclotron frequency both above and below the
ion gyroscale.
\end{abstract}


\maketitle

\paragraph{Introduction.}
A wide variety of astrophysical plasmas---e.g., the solar wind and
corona, the interstellar and intracluster medium, accretion disks
around compact objects---are magnetized and turbulent. The
turbulence in these systems is damped at small scales where the plasma
is only weakly collisional, so a kinetic description is required. It
is often a good approximation to consider small fluctuations occurring
on top of an equilibrium state with a uniform (or large-scale)
dynamically strong mean magnetic field (the Kraichnan hypothesis
\cite{Kraichnan:1965}).  The resulting (subsonic) magnetohydrodynamic
(MHD) turbulence is believed to be a Kolmogorov-like cascade of
spatially anisotropic Alfv\'enic fluctuations \cite{Goldreich:1995}.
Such anisotropy is observed in laboratory plasmas
\cite{Robinson:1971}, the solar wind \cite{Bieber:1996}, and numerical
simulations \cite{Shebalin:1983}.  Assuming a {\em critical balance}
between the linear frequencies and nonlinear decorrelation rates
\cite{Higdon:1984a,Goldreich:1995}, the anisotropy is scale-dependent
with wave numbers parallel and perpendicular to the local mean field
related by $k_\parallel \propto k_\perp^{2/3}$.  This implies that in
most astrophysical plasmas, the frequencies of the Alfv\'enic
fluctuations remain below the ion cyclotron frequency, $\omega =
k_\parallel v_A\ll \Omega_i$, even as the perpendicular wavelength
reaches the ion gyroscale, $k_\perp \rho_i \sim 1$.\\ \indent Such
fluctuations are well described by {\em gyrokinetics} (GK), a rigorous
low-frequency anisotropic limit of kinetic theory
\cite{Frieman:1982,Howes:2006,Schekochihin:2007,Howes:2007}, which
systematically averages out the cyclotron motion of particles about
the magnetic field. In GK, the MHD fast wave and cyclotron resonances
are ordered out, while finite Larmor radius effects and the
collisionless Landau resonance are retained. GK enables numerical
studies of kinetic turbulence with today's computational resources
because the gyroaveraging eliminates fast time scales and reduces the
dimensionality of phase space from six to five (three spatial
dimensions plus the parallel and perpendicular particle velocities).
GK has been used to study electrostatic turbulence in fusion plasmas
for decades, but there have been few GK treatments of astrophysical
plasma turbulence. GK is not applicable to large-scale, roughly
isotropic fluctuations, such as are directly excited in the
interstellar medium by supernovae.  However, the fluctuations in
magnetized plasma turbulence become smaller amplitude and more
anisotropic at smaller scales. GK theory and simulations are thus
appropriate, and hold considerable promise, for studies of microscopic
phenomena such as turbulent heating and magnetic reconnection, and for
interpreting observations of short-wavelength turbulent
fluctuations. This Letter reports the first \emph{ab initio}, fully
electromagnetic, kinetic simulations of turbulence in a magnetized
weakly collisional astrophysical plasma.\\ \indent The study of
turbulence in weakly collisional plasmas benefits greatly from access
to a unique laboratory---the near-Earth solar wind---in which
spacecraft make \emph{in situ} measurements of the properties of
turbulence from the large (energy-containing) scales to the small,
kinetic scales at which fluctuations are damped. The one-dimensional
frequency spectrum of magnetic fluctuations typically shows a
power-law behavior with a $-5/3$ slope at low frequencies
\cite{Goldstein:1995}, a break at a few tenths of a Hz, and a steeper
power-law at higher frequencies with a slope that varies between $-2$
and $-4$ \cite{Leamon:1998a}. It is generally agreed that the $-5/3$
range is an MHD inertial range, while the break and the
dissipation-range slope have been variously attributed to proton
cyclotron damping \cite{Coleman:1968}, Landau damping of kinetic
\Alfven waves (KAW) \cite{Leamon:1999}, or the dispersion of whistler
waves \cite{Stawicki:2001}. Recent simultaneous measurements of the
magnetic- and electric-field fluctuations found an increase in the
wave phase velocity above the spectral break \cite{Bale:2005}, a
finding consistent with the conversion to a KAW cascade but
inconsistent with cyclotron damping \cite{Howes:2007}.  The GK
simulations presented below capture all of the spectral features
described above and show magnetic- and electric-energy spectra similar
to those reported empirically in \cite{Bale:2005}.
Our simulation results suggest that the turbulent fluctuation spectra
observed in the solar wind are a consequence of the transition from an
Alfv\'en-wave to a KAW cascade.

\paragraph{The Code.} 
We have used \texttt{AstroGK}, a new GK code developed specifically to
study astrophysical turbulence.  \texttt{AstroGK} is essentially a
slab version of the publicly available code \texttt{GS2}, used to
study plasma turbulence in fusion devices \cite{Kotschenreuther:1995}.
We now give a brief overview of the code.  A detailed description will
appear elsewhere.\\ \indent The simulation domain is a periodic flux
tube with a straight uniform mean magnetic field $B_0$ and no
equilibrium gradients.  The equilibrium distribution is taken to be
Maxwellian for all particle species. The code solves the GK equation
\cite{Howes:2006}, evolving the perturbed gyroaveraged distribution
function $h_s(x,y,z,\varepsilon_s,\xi)$ of the guiding centers for
each species $s$---ions (protons) and electrons with the correct
mass ratio $m_i/m_e=1836$.  Spatial dimensions $(x,y)$ perpendicular
to the mean field are treated pseudospectrally; a conservative
finite-difference scheme is used in the parallel direction $z$. A
gyroaveraged
pitch-angle-scattering collision operator \cite{Schekochihin:2007} is
used.  The pitch-angle derivatives are done using second-order finite
differences.  The electromagnetic field is represented by the scalar
potential $\varphi$, parallel vector potential $A_\parallel$, and the
parallel magnetic field perturbation $\delta B_\parallel$.  These are
determined from the quasineutrality condition and Amp\`ere's law
\cite{Howes:2006}, where the charge densities and currents are
calculated as velocity-space moments of the perturbed distribution
function. These velocity-space integrals (over particle energies
$\varepsilon_s=m_sv^2/2$ and pitch angles $\xi=v_\parallel/v$) are
done with spectral accuracy, using high-order Gaussian--Legendre
integration rules.  The linear terms in the GK system, including the
field equations, are advanced implicitly in time; for the nonlinear
terms, an explicit, third-order Adams-Bashforth scheme is used.

\begin{figure}
\resizebox{3.in}{!}{\includegraphics*[0.25in,6.in][7.85in,9.7in]{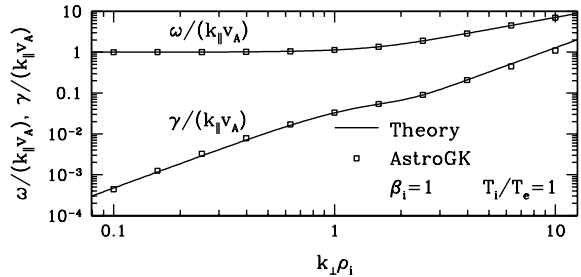}}
\vskip-0.2cm
\caption{\label{fig:linear} Normalized frequencies 
$\omega/k_\parallel v_A$ and damping rates $\gamma/k_\parallel v_A$ 
vs.\ the normalized perpendicular wave number
$k_\perp\rho_i$ for a plasma with $\beta_i=1$ and
$T_i/T_e=1$. \texttt{AstroGK} (open squares) correctly reproduces the
analytic results from the linear collisionless gyrokinetic dispersion relation
(line) \cite{Howes:2006}.}
\vskip-0.5cm
\end{figure}

\paragraph{Linear Benchmarks.} 
In an earlier paper \cite{Howes:2006}, we 
verified that \texttt{GS2} correctly describes linear 
kinetic physics in the parameter regimes relevant to astrophysical plasmas. 
\texttt{AstroGK} has been 
checked to agree with \texttt{GS2} exactly and also benchmarked against 
linear kinetic theory, as illustrated  
by \figref{fig:linear}: for $k_\perp \rho_i\ll1$, 
we have \Alfven waves, $\omega=\pm k_\parallel v_A$, and the damping is very small; 
for $k_\perp \rho_i\gg1$, these become kinetic \Alfven waves,
$\omega = \pm k_\parallel v_A k_\perp\rho_i/\sqrt{\beta_i + 2/(1+T_e/T_i)}$, 
so their phase velocity increases linearly with $k_\perp$  
and they are also more strongly damped, in excellent agreement with linear 
theory \cite{Howes:2006,Quataert:1999}. 
Here $v_A=B_0/\sqrt{4\pi m_i n_i}$ is the \Alfven speed, 
$n_i$ the ion number density, 
$T_e$ and $T_i$ the ion and electron temperatures, 
and $\beta_i=8\pi n_i T_i/B_0^2$.

\paragraph{Driving.} 
The driving and dissipation scales in astrophysical turbulence are
widely separated: e.g., in the slow solar wind, the ion gyroscale is
$\rho_i\sim10^{6}$~cm, while the effective driving scale is
$L\sim10^{11}$~cm \cite{Howes:2007}.  Such scale separations are, of
course, not accessible numerically. In our simulations, the size of
the domain is understood to be much smaller than the driving scale.
We model the energy influx from larger scales by adding to Amp\`ere's
law a parallel ``antenna'' current $j^a_{\parallel,{\bf k}}$.  For
each chosen driving wave vector $\V{k}_a$, the antenna amplitude is
calculated from a Langevin equation whose solutions are \Alfven waves
with wave vector $\V{k}_a$, frequency $\omega=\pm k_{a\parallel}v_A$
and a decorrelation rate comparable to $\omega$.  This method of
driving is motivated by the theoretical expectation that the
turbulence in the inertial range (at scales $\rho_i\ll k^{-1}\ll L$)
is Alfv\'enic and critically balanced \cite{Goldreich:1995}.

\paragraph{Dissipation.} 
The driving injects power into the system in the form of
electromagnetic fluctuations. In steady state, this power must be
dissipated into heat. By Boltzmann's H-theorem, no entropy increase
and, therefore, no heating is possible in a kinetic system without
collisions. If the collision rate is smaller than the fluctuation
frequencies,
the perturbed distribution function develops small-scale structure in
velocity space \cite{Howes:2006,Schekochihin:2007}.  This makes the
velocity derivatives in the collision integral large so the collisions
can act, a situation analogous to the emergence of small spatial
scales in neutral fluids with small viscosity (Kolmogorov cascade). In
GK turbulence, the cascades in position and velocity space are linked,
so we may speak of a kinetic cascade in five-dimensional phase space
\cite{Schekochihin:2007}. Collisionless Landau damping of the
electromagnetic fluctuations leads to particle heating in the sense
that it transfers the electromagnetic fluctuation energy into
fluctuations of the particle distribution function (the kinetic
entropy cascade \cite{Schekochihin:2007}), which are then converted
into heat by collisions.\\ \indent A detailed analysis of the kinetic
cascade will be presented in a separate study, but the lesson is that
kinetic turbulence simulations need to include collisions and need to
have sufficient velocity-space resolution for the correct relationship
to be established between small-scale structures in velocity and
position space.  Accomplishing this with a physical collision operator
simultaneously for ions and electrons is very difficult. To ease the
resolution requirements, we employ a hypercollisionality (analogous to
hyperviscosity in fluid simulations). This takes the form of a
pitch-angle-scattering operator with a wave-number-dependent collision
rate $\nu_{h} (k_\perp/k_{\perp\rm max})^8$, where $k_{\perp\rm max }$
is the grid-scale wave number.  This artificially enhanced collision
term terminates the cascade and produces positive-definite heating
close to the grid scale, while allowing essentially collisionless
physics at larger scales.  For the ions, the importance of the
hypercollisionality is marginal, while for the electrons, we needed a
large value of $\nu_{h}$.  As a result, electron heating (at the
electron gyroscale $\rho_e$) is not well modeled, but this is an
acceptable sacrifice because our focus is on the turbulent cascade
through the ion gyroscale at $\rho_i\gg\rho_e$.

\begin{figure}[t!]
\resizebox{3.in}{!}{\includegraphics*[0.25in,2.in][7.85in,5.75in]{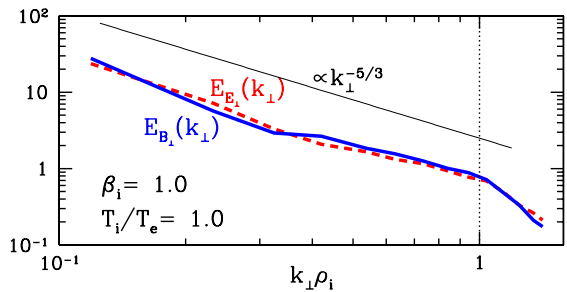}}
\caption{\label{fig:swm_spec}(Color online) Magnetic (solid line) and electric
(dashed line) energy spectra in the MHD regime ($k_\perp\rho_i<1$).
The box size is $L_\perp/2\pi=10\rho_i$. Electron hypercollisionality
is dominant for $k_\perp\rho_i \ge 1$ (dotted line).}
\vskip-0.5cm
\end{figure}

\paragraph{Results.} 
The physical parameters in GK simulations of plasma turbulence are
$T_i/T_e$ and $\beta_i$.  Here both are set to~1, sensible
characteristic values for the solar wind at 1 AU, and for the
interstellar medium; a full parameter scan is clearly desirable in the
future (e.g., $\beta_i$ in the solar wind at 1 AU varies roughly
between $0.1$ and $10$).  By varying the driving wave number $k_a$ and
the (hyper)collision rate, we may focus on various scale ranges. Here
we present results obtained for the inertial range
($k_\perp\rho_i\ll1$) and around the ion gyroscale
($k_\perp\rho_i\sim1$).  
In what follows, the normalized magnetic-energy spectrum is defined
$E_{B_\perp}(k_\perp) = (L_z/L_\perp^2) 2\pi k_\perp^3\int dz\,
{\langle|A_{\parallel,\V{k}_\perp}(z)|^2\rangle/8\pi n_i T_i}$,
where $k_\perp$ is measured in units of $\rho_i^{-1}$,
$L_z$ and $L_\perp$ are parallel and perpendicular box
dimensions, and the angle brackets denote angle averaging
over a wavenumber shell centered at $|\V{k}_\perp|=k_\perp$
and with the width equal to $2\pi/L_\perp$.
The normalized electric-energy spectrum $E_{E_\perp}(k_\perp)$ is
defined in a similar way in terms of $\varphi_{\V{k}_\perp}$, with an
extra factor of $(c/v_A)^2$, where $c$ is the speed of light.\\
\indent In the inertial range, $k_\perp\rho_i\ll1$, the Reduced MHD
equations are the rigorous limit of GK for Alfv\'enic fluctuations
\cite{Schekochihin:2007}.  Thus, kinetic turbulence in this regime
must be consistent with the numerical results obtained in MHD
simulations \cite{Shebalin:1983}.
\figref{fig:swm_spec} shows the normalized magnetic and electric
energy spectra calculated gyrokinetically in this regime. As expected
for critically balanced Alfv\'enic turbulence \cite{Goldreich:1995},
these spectra are coincident and have a scaling consistent with
$k_\perp^{-5/3}$.  This is the first demonstration of an MHD
turbulence spectrum in a {\em kinetic} simulation.  While this is not
a surprising result, it can be viewed as a fully nonlinear
benchmark.\\
\begin{figure}[t!]
\resizebox{3.in}{!}{\includegraphics*[0.25in,2.in][7.85in,7.7in]{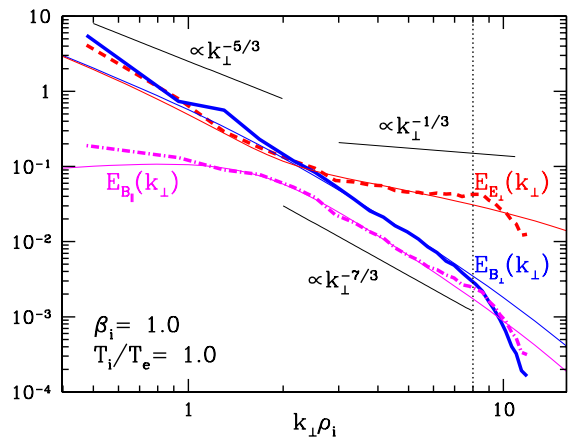}}
\caption{\label{fig:swt_spec}(Color online) Bold lines: normalized energy spectra
for $\delta B_\perp$ (solid), $\delta B_\parallel$ (dash-dotted), and
$E_\perp$ (dashed). Thin lines: solution of the turbulent cascade
model of \cite{Howes:2007}.  The resolution of this simulation is
$(N_x,N_y,N_z,N_{\varepsilon},N_{\xi},N_s) =(64,64,128,8,64,2)$,
requiring $\simeq 0.5 \times 10^9$ computational mesh points.  The box
size is $L_\perp/2\pi=2.5\rho_i$. Electron hypercollisionality is
dominant for $k_\perp\rho_i\ge8$ (dotted line). }
\vskip-0.5cm
\end{figure}
\indent Our main numerical experiment focuses on scales near $k_\perp
\rho_i\sim 1$.  This regime cannot be simulated by any fluid
model. However, we know from theory that low-frequency Alfv\'enic
turbulence is rigorously described by Reduced MHD equations for
$k_\perp\rho_i\ll 1$ and by a similarly reduced version of the
Electron MHD equations for $k_\perp\rho_i\gg 1$
\cite{Schekochihin:2007}.  The latter system supports kinetic Alfv\'en
waves (see \figref{fig:linear}).  If one assumes a turbulent cascade
of KAW-like fluctuations decorrelating on a timescale comparable to
the linear KAW period (critical balance), a Kolmogorov-style scaling
argument predicts that the magnetic-energy spectrum steepens from
$k_\perp^{-5/3}$ to $k_\perp^{-7/3}$, while the electric-energy
spectrum flattens to $k_\perp^{-1/3}$
\cite{Biskamp:1999,Schekochihin:2007,Howes:2007}.  Thus, a spectral
break is expected around $k_\perp\rho_i\sim1$, corresponding to the
transition between Alfv\'en-wave and KAW turbulence.
\figref{fig:swt_spec} shows the energy spectra in our simulations near
this transition.
A spectral break is, indeed, observed (at $k_\perp \rho_i \simeq 2$),
as is the steepening (flattening) of the magnetic-(electric-)energy
spectra.  The spectra at wave numbers below and above the transition
are consistent with the above predictions for critically balanced
Alfv\'en-wave and KAW cascades
\cite{Goldreich:1995,Biskamp:1999,Schekochihin:2007,Howes:2007}.\\
\indent There is a striking similarity between the simulated spectra
shown in \figref{fig:swt_spec} and the magnetic- and electric-energy
spectra in the solar wind reported in \cite{Bale:2005}.
The increase in phase velocity in the dissipation range
($k_\perp\rho_i>1$), shown by both measurement and simulation, is
compelling evidence that the observed breaks in the spectra are caused
by a transition to a KAW cascade, not by the onset of ion cyclotron
damping \cite{Howes:2007}.\\ \indent The scaling predictions for KAW
turbulence are made assuming negligible Landau damping. In our
simulations, the damping is, indeed, small, so it is reasonable that
the scaling predictions are well satisfied. However, this will not be
true in all real astrophysical situations. We have argued
\cite{Howes:2007} that the spectra much steeper than $k_\perp^{-7/3}$
often observed in the solar wind \cite{Leamon:1998a} can be due to
non-negligible Landau damping.  A simple way to estimate the effects
of the damping on the energy spectra was proposed in \cite{Howes:2007}
(see also \cite{Quataert:1999}), where a spectral model of the
turbulent cascade was constructed based on three assumptions: (i)
spectrally local energy transfer, (ii) critical balance, (iii) the
applicability of the linear damping rates.
Using this model, the energy spectrum $E_{B_\perp}(k_\perp)$ can be
predicted in the entire simulation range, given one ``Kolmogorov''
constant, which quantifies the linear damping rate relative to the
non-linear cascade rate. In \figref{fig:swt_spec}, we show that this
analytical model reproduces the entire shape of the numerical
spectrum.  The model works well without fine tuning, for a range of
values of the constant; this is because the damping is small in this
simulation and our model captures the transition from Alfv\'enic to
KAW turbulence.  The agreement between the analytical model and the
simulations is a non-trivial result: it suggests that the linear
damping rate does not significantly underestimate the rate at which
the electromagnetic energy is dissipated in the nonlinear simulations.
Future simulations will determine whether stronger linear damping can
account for the steeper spectra often observed in the solar wind.\\
\indent A further test of the conclusion that we are seeing a KAW
cascade in the simulations is achieved by using the linear GK
eigenfunctions for KAWs to produce the energy spectra for the electric
fluctuations ($E_\perp$) and for the fluctuations of the field
strength ($\delta B_\parallel$). These fit the spectra measured from
our numerical data well (\figref{fig:swt_spec}).

\paragraph{Conclusions.} 
We have presented first-of-a-kind kinetic simulations of turbulence in
a weakly collisional, magnetized plasma. The ion-gyroscale turbulent
fluctuations simulated here represent the fate of a larger-scale MHD
cascade.  The qualitative agreement between our simulations and
solar-wind measurements \cite{Bale:2005} supports theoretical models
in which the turbulent fluctuations in the solar wind have frequencies
well below the ion cyclotron frequency even when the cascade reaches
the (perpendicular) scale of the ion Larmor radius.  The observed
break in the magnetic-energy spectrum in the solar wind is inferred to
correspond to a transition to kinetic-Alfv\'en-wave turbulence, not to
the onset of ion cyclotron damping.  Although half a billion
mesh points were used in the case of \figref{fig:swt_spec}, the
resolution in velocity space is still not fully sufficient to draw
detailed conclusions about the turbulent heating. Nonetheless, the
agreement between the simulations and an analytical cascade model
based on linear damping rates implies that the latter do not
significantly underestimate the true damping in a turbulent
collisionless plasma.  Future simulations will probe a range of plasma
parameters, including more heavily damped regimes, that will allow a
more quantitative study of the role of collisionless damping in
turbulent plasmas.
The first results reported in this Letter demonstrate that such kinetic 
simulations of plasma turbulence may be undertaken with some confidence, 
using existing computational resources.  

\begin{acknowledgments}
This work was supported by the US DOE Center for Multiscale Plasma 
Dynamics (G.G.H., T.T.), 
the David and Lucille Packard Foundation (E.Q.), 
UK STFC (A.A.S.), the Leverhulme Trust  
Network for Magnetized Plasma Turbulence 
and the Aspen Center for Physics. 
\end{acknowledgments}


\end{document}